\documentclass[10pt,conference]{IEEEtran}
\usepackage{amsmath}
\usepackage{amssymb,palatino,theorem}
\usepackage{graphicx}
\usepackage{cite}

\newcommand{\spazio}{\rule[-6mm]{0mm}{0mm}}

\newcommand{\RE}{{\rm Re}}
\newcommand{\IM}{{\rm Im}}
\def\xv     {\mbox{\boldmath $x$}}
\def\av     {\mbox{\boldmath $a$}}
\def\dv     {\mbox{\boldmath $d$}}
\def\uv     {\mbox{\boldmath $u$}}
\def\yv     {\mbox{\boldmath $y$}}
\def\zv     {\mbox{\boldmath $z$}}

\def\Hm     {\mbox{\boldmath $H$}}
\def\Xm     {\mbox{\boldmath $X$}}
\def\Ym     {\mbox{\boldmath $Y$}}
\def\Zm     {\mbox{\boldmath $Z$}}
\def\Em     {\mbox{\boldmath $E$}}
\def\Am     {\mbox{\boldmath $A$}}
\def\Nm     {\mbox{\boldmath $N$}}
\def\Pm     {\mbox{\boldmath $P$}}

\def\Idm    {\mbox{\boldmath $I$}}
\def\Um     {\mbox{\boldmath $U$}}
\def\Vm     {\mbox{\boldmath $V$}}
\def\Sm     {\mbox{\boldmath $\Sigma$}}
\def\Lm     {\mbox{\boldmath $\Lambda$}}
\def\Lv     {\mbox{\boldmath $\lambda$}}

\def\Dm     {\mbox{\boldmath $D$}}

\def\xv     {\mbox{\boldmath $x$}}
\def\yv     {\mbox{\boldmath $y$}}

\def\Hm     {\mbox{\boldmath $H$}}
\def\Xm     {\mbox{\boldmath $X$}}
\def\Ym     {\mbox{\boldmath $Y$}}
\def\Zm     {\mbox{\boldmath $Z$}}
\def\Nm     {\mbox{\boldmath $N$}}
\def\Pm     {\mbox{\boldmath $P$}}

\def\Idm    {\mbox{\boldmath $I$}}

\newtheorem{proposition}{Proposition}

\newtheorem{example}{Example}

\makeatletter
\def\@begintheorem#1#2{\it \trivlist \item[\hskip \labelsep{\bf #1\
#2.}]} \makeatother

\newcommand{\be}{\begin{equation}}
\newcommand{\ee}{\end{equation}}
\newcommand{\bea}{\begin{eqnarray}}
\newcommand{\eea}{\end{eqnarray}}

\newcommand{\beq}[1]{\begin{equation}\label{#1}}
\newcommand{\eeq}{\end{equation}}

\newcommand{\beqn}[1]{\begin{eqnarray}\label{#1}}
\newcommand{\eeqn}{\end{eqnarray}}

\newcommand{\beaa}{\begin{eqnarray*}}
\newcommand{\eeaa}{\end{eqnarray*}}
\def\E{{\mathbb E}}

\def\SNR    {\mbox{\scriptsize\sf SNR}}

\def \det       {{\rm det}}

\def \nT        {n_\mathrm{\scriptscriptstyle T}}
\def \nR        {n_\mathrm{\scriptscriptstyle R}}
\def \nb        {n_\mathsf{\scriptstyle b}}
\def \np        {n_\mathsf{\scriptstyle p}}

\def \tr        {\mathrm{Tr}}

\def\AbvGT #1#2{\lower2pt\vbox{\baselineskip0pt \lineskip-.5pt%
         \halign{$#1 ##$\cr #2\crcr >\cr}}}

\def\complex{\mathop{\raise .45ex\hbox{${\bf\scriptstyle{|}}$}
      \kern -0.40em {\rm \textstyle{C}}}\nolimits}
\def\hilbert{\mathop{\raise .21ex\hbox{$\bigcirc$}}\kern -1.005em
{\rm\textstyle{H}}} 

\begin{document}

\title{Mutual Information of IID Complex Gaussian Signals on Block Rayleigh-faded Channels}

\author{
\authorblockN{Fredrik Rusek}
\authorblockA{Lund University\\
221 00 Lund, Sweden\\
Email: fredrik.rusek@eit.lth.se}
\and
\authorblockN{Angel Lozano}
\authorblockA{
Universitat Pompeu Fabra\\
Barcelona 08003, Spain\\
Email: angel.lozano@upf.edu}
\and
\authorblockN{Nihar Jindal}
\authorblockA{University of Minnesota\\
Minneapolis, MN 55455\\
Email: nihar@umn.edu}
}

%

\maketitle

\begin{abstract}
We present a method to compute, quickly and efficiently, the mutual information achieved by an IID (independent identically distributed) complex Gaussian input
on a block Rayleigh-faded channel without side information at the receiver. The method accommodates both scalar and MIMO (multiple-input multiple-output) settings.
Operationally, the mutual information thus computed represents the highest spectral efficiency that can be attained using standard Gaussian codebooks.
Examples are provided that
illustrate the loss in spectral efficiency caused by fast fading and how that loss is amplified by the use of multiple transmit antennas.
These examples are further enriched by comparisons with the channel capacity under perfect channel-state information at the receiver, and with the spectral
efficiency attained by pilot-based transmission.

\end{abstract}

\section{Introduction}

IID (independent identically distributed) complex Gaussian inputs are most relevant in wireless communication impaired by Gaussian noise.
Some of the arguments for this relevance are that, if perfect CSI (channel state information) is available at the receiver, then:
\begin{itemize}
\item These are the unique capacity-achieving inputs.
\item Their mutual information represents very well the mutual information of proper complex discrete constellations (e.g., QAM) commonly used in wireless systems.
(This holds up to some power level that depends on the cardinality of the constellation \cite{mercury_journal}.)
\end{itemize}
Expressions for the perfect-CSI capacity achieved by IID complex Gaussian inputs are available (cf. Section \ref{baselines}) and thus such capacity can be easily evaluated.

Remove now the perfect CSI.
No expressions are available for the mutual information achieved by IID complex Gaussian inputs save for the special case of memoryless channels \cite{perera07}.
Moreover, straight Monte-Carlo computation is not feasible because it would entail large-dimensional histograms.
Only asymptotic expansions of the mutual information in the low- and high-power regimes \cite{Zheng02,prelov,RaoHassibi,lapidoth05} have been derived.
And yet, although no longer capacity-achieving \cite{marzetta2,ShamaiTrott}, IID complex Gaussian inputs remain highly relevant.
Operationally, the mutual information they achieve represents the highest spectral efficiency that can be attained using standard Gaussian codebooks.
In fact, the capacity-achieving inputs in the absence of perfect-CSI become largely impractical in many cases
(in the low-power regime, for example, they become unacceptably peaky)
and thus the capacity is arguably less relevant to system designers than the mutual information of IID complex Gaussian inputs.

This paper presents analytical expressions for the output distributions of a block Rayleigh-faded channel fed with IID complex Gaussian inputs.
Then, a simple outer Monte-Carlo in conjunction with these distributions yields a semi-analytical method that allows
evaluating, quickly and efficiently, the mutual information. 
The method accommodates not only scalar channels, but also MIMO (multiple-input multiple-output) settings.
Altogether, this allows answering questions such as:
\begin{itemize}
\item What is the impact of the perfect-CSI assumption?
\item How suboptimal are pilot-based schemes, i.e., those schemes that form explicit channel estimates on the basis of pilot observations at the receiver and subsequently
apply these estimates to detect the data?
\item At which power level is the power efficiency maximized (i.e., the energy per bit minimized)?
\end{itemize}
Not surprisingly, the answers to these questions end up being a function of the fading rate and the numbers of antennas. With the method presented, these
relationships can be precisely established, and some examples of this are provided in the paper.

\section{Channel Model}

Consider $\nT$ transmit and $\nR$ receive antennas, with $\nT \geq \nR$, and
let the $\nR \times \nT$ matrix $\Hm$ represent the discrete-time fading channel.
Under block Rayleigh-fading, the channel entries are drawn from a zero-mean unit-variance complex Gaussian distribution at the beginning of each
fading block and they then remain constant for the $\nb$ symbols within that block, where $\nb$
is the coherence time in symbols (or the coherence bandwidth if it is the frequency domain being modeled).
This process is repeated for every block in an IID fashion.
There is no antenna correlation and thus, within a given fading block, the entries of $\Hm$ are also independent.

Assembling on matrices the input, the output, and the noise for the $\nb$ symbols within each block,
their relationship becomes
\be
\label{ibrahimovich}
\Ym  =  \sqrt{\frac{\SNR}{\nT}} \Hm \Xm + \Nm
\ee
where the input $\Xm$ is an $\nT \times \nb$ matrix while the output $\Ym$ and the noise $\Nm$ are $\nR \times \nb$ matrices. Both $\Xm$ and $\Nm$ have
IID zero-mean unit-variance complex Gaussian entries. With that, $\SNR$ indicates the average signal-to-noise ratio per receive antenna.

Each codeword spans a large number of fading blocks, in the time and/or frequency domains,
which endows ergodic quantities with operational meaning.

Although a block-fading structure admittedly represents a drastic simplification of reality, it does capture the essential nature of fading and yields results that are
remarkably similar to those obtained with continuous-fading models \cite{lozano08}.
In fact, for a rectangular Doppler spectrum, an exact correspondence in terms of the error in the estimation of $\Hm$
 can be established between block- and continuous-fading models \cite{JindalLozano09} whereby
\be
\label{busquets}
\nb = \frac{1}{2 f_{\sf m} T_{\rm s}}
\ee
where $f_{\sf m}$ and $T_{\rm s}$ are the maximum Doppler frequency and the symbol period, respectively.
Typically, $f_{\sf m} = (v/c) f_{\sf c}$ with $v$ the velocity and $f_{\sf c}$ the carrier frequency.
The mapping in (\ref{busquets}) is in terms of the minimum mean-square error in the estimation of $\Hm$,
and thus it is exact for pilot-based schemes that rely on such explicit estimation,
but more broadly we take it as indicative of the fading rate represented by a given value of $\nb$.

\section{Computation of the Mutual Information}
\label{method}

The mutual information under investigation can be expressed as
\be
\label{alves}
\bar{\mathcal{I}}=\frac{1}{\nb}\left[\mathfrak{h}(\Ym)-\mathfrak{h}(\Ym|\Xm)\right]
\ee
where $\mathfrak{h}(\cdot)$ denotes the differential entropy operator.

Our first result leverages the derivations in \cite{shinlee} to obtain a closed-form expression for $\mathfrak{h}(\Ym|\Xm)$.

\begin{proposition}
\label{R0}
Let  $E_q(\cdot)$ denote the exponential integral
\be
\label{exponentialintegral}
E_q(\zeta)=\int_1^{\infty} t^{-q} e^{-\zeta t} dt .
\ee
Then,
\begin{align}
\mathfrak{h}(\Ym|\Xm) =& \; \nR \log_2(e) \; e^{\nT/\SNR} \sum_{i=0}^{\nT-1} \sum_{j=0}^i \sum_{\ell=0}^{2j}
\left[
\left( \! \begin{array}{c}
                2i-2j \\
                i-j
              \end{array} \! \right)
\right. \nonumber \\
& \cdot \left( \! \begin{array}{c}
                2j+2 \nb - 2 \nT \\
                2j-\ell
              \end{array} \!\! \right) \frac{(-1)^{\ell} \, (2j)! \, (\nb-\nT+\ell)!}{2^{2i-\ell} \, j! \, \ell! \, (\nb-\nT+j)!} \nonumber \\
& \left. \sum_{q=0}^{\nb-\nT+\ell} E_{q+1} \! \left( \frac{\nT}{\SNR} \right)
\right] + \nR \nb \log_2(\pi e)
\label{siscopes}
\end{align}

\vspace*{2mm}
\noindent {\bf Proof:} See Appendix.
\end{proposition}




Having expressed $\mathfrak{h}(\Ym|\Xm)$ in (\ref{alves}), we now turn our attention to $\mathfrak{h}(\Ym)$.
Unfortunately, a closed-form solution for it seems out of reach.
Denoting by $p(\Ym)$ the unconditional output distribution,
\begin{align}
\mathfrak{h}(\Ym)&=-\int p(\Ym)\log_2 p(\Ym) \mathrm{d}\Ym \\
&= -\E \left[ \log_2 p(\Ym)\right]
\end{align}
Thus, if a formula for $p(\Ym)$ were at hand, $\mathfrak{h}(\Ym)$ could be
evaluated by straightforward Monte Carlo averaging.
Such a formula is the object of the next result. 

\begin{proposition}
\label{R1}
For $1\leq k \leq \nR$, define the functions
\be
f_k(x) \! = \!\! \int_0^{\infty}\!\!\!\!\exp\! \left\{ \!\frac{x\,\SNR\,
    z}{z\SNR\!+\!\nT}\!-\!z\! \right\}
\!\frac{z^{k-1+\nR-\nT}\,(\nT/\SNR)^{k-1}}{\left(z\SNR/\nT\!+\!1\right)^{  {n_\mathsf{\scriptscriptstyle b}} +1-\nR}}\mathrm{d}z
\ee
Let $\dv = [d_1,\ldots,d_{\nR}]$ be the eigenvalues of $\Ym^\dagger \Ym$ and
    define the $\nR\times \nR$ matrix $\Zm$ as $Z_{ij}=f_i(d_j), \, 1\leq i,j \leq \nR.$
Then   
\be
\label{Py}
p(\Ym)\:=\:\frac{K(\Ym)}{\prod_{1 \leq i,j \leq \nR} (d_j - d_i) \prod_{k=0}^{\nR-1} (\nT-k-1)! } \, \det \Zm
\ee
where
\be
\label{Ky}
K(\Ym) = \frac{e^{-  \|  {\mbox{\boldmath \scriptsize $Y$}} \|^2 }}{\pi^{ {n_\mathsf{\scriptscriptstyle b}} \nR }} .
\ee

\vspace*{2mm}
\noindent {\bf Proof:} See Appendix.
\end{proposition}

In the special case of memoryless channels, i.e., for $\nb=1$, the solution in Proposition \ref{R1} reduces to the one in  \cite{perera07}.

We also note that, due to the rotational invariance of $\Hm$, the phase of the entries of $\Ym$ carries no information about $\Xm$.
Only the magnitude of the entries of $\Ym$ is relevant to the distribution in (\ref{Py}).


Using (\ref{siscopes}) and (\ref{Py}), an algorithm to compute $\bar{\mathcal{I}}$ can be put forth as follows.
\vspace*{0.5mm}
\begin{center}
\begin{tabular}{|p{80mm}|}
\hline
\tt{Algorithm 1:Evaluation of $\bar{\mathcal{I}}$.} \\
\hline \hline
1. Pre-compute $f_k(x)$, $1\leq k \leq \nR$, on a discrete set
$\mathcal{X}$ with a suitable stepsize $\Delta x = x_k-x_{k-1}$. \\ \hline
2. Generate a sufficiently large number of input and output vectors according to (\ref{ibrahimovich}).
\\ \hline
3. For each input and output pair, apply (\ref{Py}) to obtain $p(\Ym)$.
\\ \hline
4. Compute the sample average of $-\log_2p(\Ym)$ via Monte Carlo, thereby obtaining $\mathfrak{h}(\Ym)$.
\\ \hline
5. Compute $\mathfrak{h}(\Ym|\Xm)$ from (\ref{siscopes}) and apply (\ref{alves}).
\\ \hline
\end{tabular}
\end{center}
\vspace*{0.5mm}

The accuracy can be made as high as desired by averaging over more input/output sample pairs
and by increasing the precision in Step 1.
For the results presented in Section \ref{examples}, the number of samples and the value of $\Delta x$ were chosen
such that two decimal digits are correct with $90\%$ probability.
With a standard workstation, the entire computation process is a matter of seconds.

\section{Baselines}
\label{baselines}

Before exemplifying the method described in Section \ref{method}, and in order to use them as baselines,
we introduce the perfect-CSI capacity and the pilot-based communication boundaries.

\subsection{Capacity with Perfect CSI}

If the receiver is provided with perfect CSI, the ergodic capacity, in bits/s/Hz, equals
\be
C(\SNR) = \E \left[ \log_2 \det \left( \Idm + \frac{\SNR}{\nT} \, \Hm \Hm^\dagger \right) \right]
\label{CapMIMO}
\ee
a closed form for which can be found in \cite{shinlee}.
For $\nT = \nR = 1$, this closed form reduces to
\be
\label{CapSISO}
C(\SNR) = e^{1/\SNR} E_1\! \left( \frac{1}{\SNR} \right) \log_2 e
\ee
where $E_1(\cdot)$ is the exponential integral
\be
E_1(\xi) = \int_1^\infty t^{-1} e^{-\xi t} dt .
\ee

At high $\SNR$, the slope of $C(\SNR)$ is $S_\infty = \min\{ \nT,\nR \}$ bits/s/Hz/(3 dB).

%

\subsection{Pilot-Based Communication}

In pilot-based communication, $\np$ pilot symbols are inserted within each fading block, leaving $\nb-\np$ symbols available for data transmission.
The channel is estimated on the basis of the pilot observations at the receiver, and this estimate is subsequently utilized to detect the data.

During the transmission of pilot symbols,
\be
\Ym_{\sf p} = \sqrt{\frac{\SNR}{\nT}} \Hm \Pm + \Nm_{\sf p}
\ee
where the output, $\Ym_{\sf p}$, and the noise, $\Nm_{\sf p}$, are $\nR \times \np$ matrices.
The entries of $\Nm_{\sf p}$ are IID zero-mean unit-variance complex Gaussian
while $\Pm$ is deterministic and must satisfy $\Pm \Pm^{\dagger} = \np \Idm$ \cite{hassibi}.

During the transmission of data symbols, in turn, (\ref{ibrahimovich}) applies only with
$\Xm$ and $\Nm$ of dimension $\nT \times (\nb-\np)$ and $\nR \times (\nb-\np)$, respectively.

The value of $\np$, which can be optimized by solving a convex problem, depends on $\SNR$, $\nb$ and $\nT$.
This optimization, and the ensuing spectral efficiency, has been studied extensively, e.g.,
\cite{hassibi,medard,Zheng02,MaYangGiannakis,Baltersee01,Ohno02,Furrer07,DengHaimovich07}.
In bits/s/Hz, such spectral efficiency applying the channel estimate as if it were the true channel \cite{amos-shamai} equals
\be
\label{aruba}
\max_{\np: 1 \leq \np \leq \nb} \left( 1-\frac{\np}{\nb} \right) \, C \! \left( \frac{\SNR^2 \np / \nT}{1+\SNR \, (1+\np / \nT)} \right)
\ee
where $C(\cdot)$ is the perfect-CSI capacity in (\ref{CapMIMO}) and (\ref{CapSISO}). A more elaborate receiver that decoded on the basis of the joint distribution of the channel
estimate and the true channel, rather than by assuming that the channel estimate equals the true channel, would exceed (\ref{aruba}) but only slightly \cite{lozano08}.

If the pilot and data symbols are not required to have the same power, i.e., if pilot power-boosting is allowed, then it is optimal to set $\np = \nT$ and to optimize
only over the relative powers of pilots and data. This results in a different convex optimization, which in this case can be solved explicitly \cite{hassibi}
leading to\footnote{Eq. (\ref{austin}) requires that $\nb > 2 \, \nT$; variations thereof are also available for $\nb \leq 2 \, \nT$ \cite{hassibi}.}
\be
\label{austin}
\left( 1-\frac{\nT}{\nb} \right) \, C \! \left( \frac{\nb \SNR}{\nb-2 \, \nT} \left( \sqrt{\gamma} - \sqrt{\gamma-1} \right) ^2 \right)
\ee
in bits/s/Hz, and with
\be
\gamma = \frac{\nb \SNR + \nT}{\nb \SNR \, \frac{{n_\mathsf{\scriptscriptstyle b}} - 2 \, \nT}{{n_\mathsf{\scriptscriptstyle b}} -\nT}} .
\ee
The spectral efficiency in (\ref{austin}) is superior to that in (\ref{aruba}).
However, pilot power boosting increases the peakiness of the overall signal distribution, rendering it less amenable to efficient amplification.

It can be easily verified that the high-$\SNR$ slope of $\bar{\mathcal{I}}_{\sf pb} (\SNR)$ equals
\be
\label{mourinho}
 \min \{ \nT, \nR \} \left( 1- \frac{\nT}{\nb} \right)
\ee
bits/s/Hz/(3 dB), which for $\nT \leq \nR$ coincides with the high-$\SNR$ slope of the channel capacity without CSI at the receiver \cite{Zheng02}:
 \be
 \label{inter}
 \min \{ \nT, \nR \} \left( 1- \frac{\min \{ \nT, \nR \}}{\nb} \right) .
 \ee
For $\nT > \nR$, (\ref{mourinho}) can also be made to equal (\ref{inter}) by using only a subset $\nR$ of the transmit antennas.
Thus, pilot-based communication need not a loss in high-$\SNR$ slope as long as pilot power boosting is allowed.
Furthermore, it follows that the high-$\SNR$ slope of $\bar{\mathcal{I}}(\SNR)$ also coincides with (\ref{inter}) and that, at high $\SNR$, it is undesirable to activate
$\nT > \nR$ transmit antennas. In the next section we shall examine whether this holds in a wider range of $\SNR$.

\section{Some Examples}
\label{examples}

In order to calibrate the relevant values of $\nb$, the following observations can be made in the context of
emerging systems such as 3GPP LTE \cite{lte} or IEEE 802.16 WiMAX \cite{wimax}:
\begin{itemize}
\item The carrier frequency $f_c$ lies between  $1$ and $5$ GHz.
\item The symbol period is $T_{\rm s} \approx 100$ $\mu$s. However, it could be
shortened to $T_{\rm s} \approx 10$-$20$ $\mu$s and the flat-faded model in (\ref{ibrahimovich}) would still apply.
(For wider bandwidths, a frequency-selective model would be required and the computation algorithm would have to be modified accordingly.)
\item Velocities up to $v \approx 120$ Km/h are of interest, and for high-speed trains this extends to $v \approx 300$ Km/h.
\end{itemize}
With all of this taken into account, $\nb$ can take values ranging from just over unity to several hundred.
As the following example evidences, for large $\nb$ the perfect-CSI capacity accurately represents the achievable mutual information.

\begin{figure}
  \begin{center}
  \includegraphics[width=3.5in]{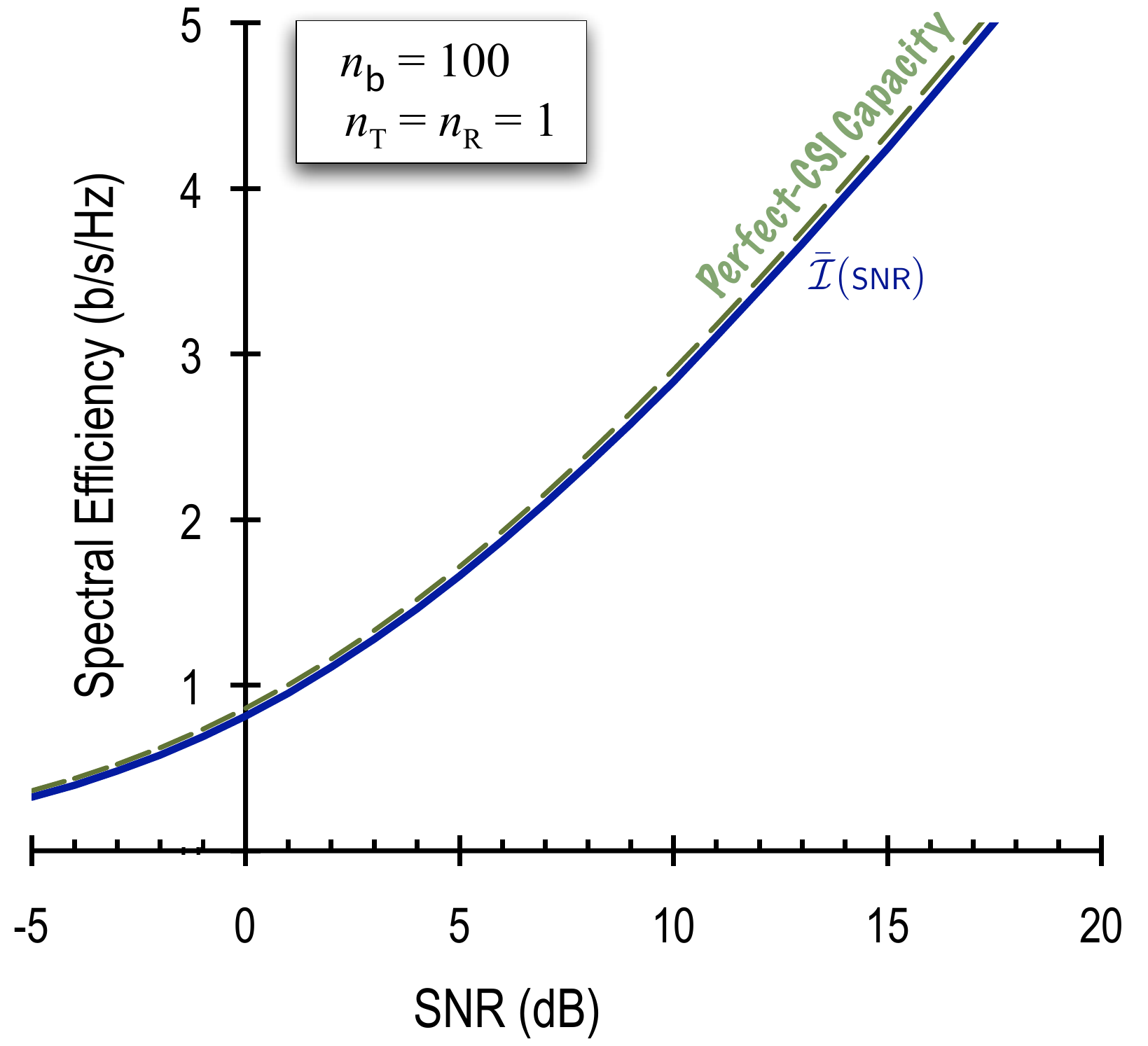}
     \end{center}
  \caption{In solid, mutual information $\bar{\mathcal{I}}(\SNR)$ for $\nT=\nR=1$ with $\nb=100$. In dashed, the perfect-CSI capacity.}
  \label{MI_SISO_100}
\end{figure}

\begin{example}
Let $\nT=\nR=1$ and let $\nb=100$. Shown in Fig. \ref{MI_SISO_100} are the mutual information and the perfect-CSI capacity as function of $\SNR$.
\end{example}

For the remainder of this section, we shall focus on scenarios
where $\nb$ is small. Specifically, we shall use $\nb=10$ and $\nb=4$.
These will tend to correspond to vehicular and high-speed-train velocities, possibly in conjunction with relatively long symbol periods and relatively high
carrier frequencies.


\begin{figure}
  \begin{center}
  \includegraphics[width=3.5in]{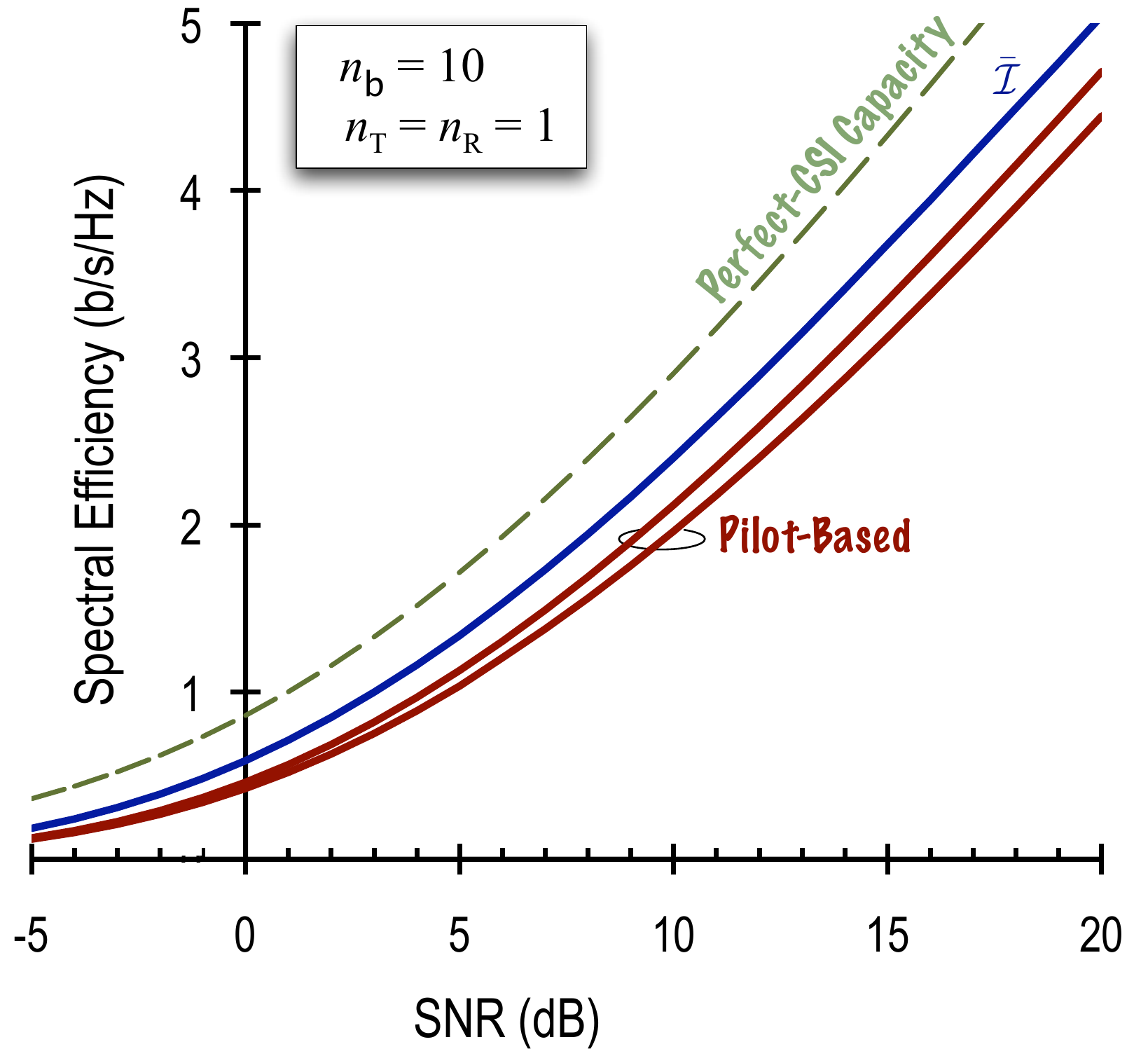}
     \end{center}
  \caption{In solid, mutual information $\bar{\mathcal{I}}(\SNR)$ for $\nT=\nR=1$ with $\nb=10$. Also in solid, spectral efficiencies achieved by pilot-based  communication, with and without pilot power boosting. In dashed, the perfect-CSI capacity.}
  \label{MI_SISO_10}
\end{figure}

\begin{example}
\label{etoo}
Let $\nT=\nR=1$ and let $\nb=10$. Shown in Fig. \ref{MI_SISO_10} is the mutual information as function of $\SNR$.
Also shown are the spectral efficiencies achieved by pilot-based communication, with and without pilot power boosting, and the perfect-CSI capacity.
\end{example}

We observe that a hefty share of the perfect-CSI capacity is achieved at high $\SNR$ in this scenario,
although this diminishes markedly with the $\SNR$.
We further observe that, by optimizing the pilot overhead or the pilot power boost at every $\SNR$, pilot-based
communication schemes can perform remarkably close to the fundamental communication limit of IID complex Gaussian inputs in this scenario.

\begin{figure}
  \begin{center}
  \includegraphics[width=3.5in]{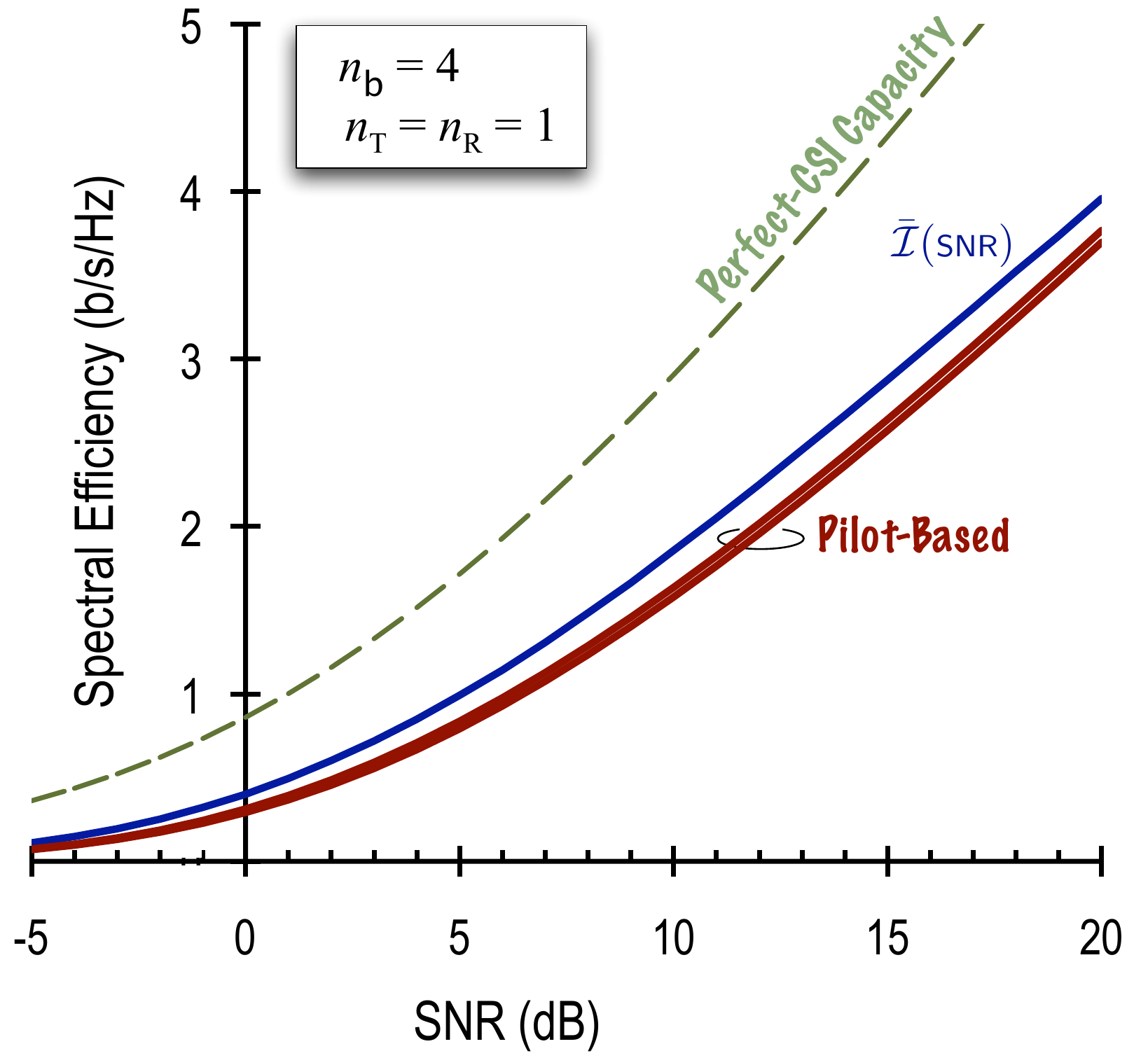}
     \end{center}
  \caption{In solid, mutual information $\bar{\mathcal{I}}(\SNR)$ for $\nT=\nR=1$ with $\nb=4$. Also in solid, spectral efficiencies achieved by pilot-based  communication, with and without pilot power boosting. In dashed, the perfect-CSI capacity.}
  \label{MI_SISO_4}
\end{figure}

\begin{example}
\label{etoo2}
Fig. \ref{MI_SISO_4} re-evaluates Example \ref{etoo} with $\nb=4$.
\end{example}

\begin{figure}
  \begin{center}
  \includegraphics[width=3.5in]{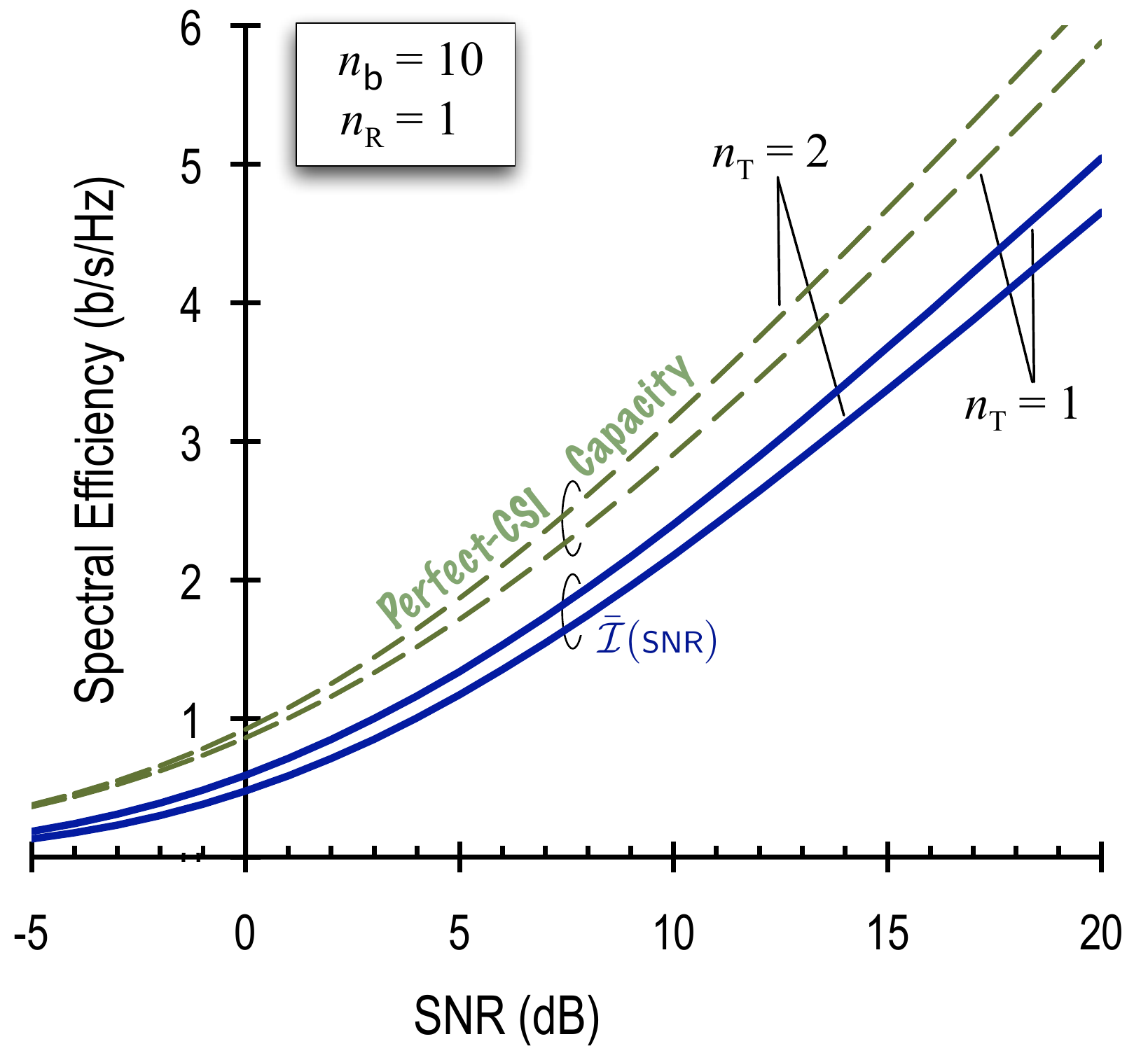}
     \end{center}
  \caption{In solid, mutual information $\bar{\mathcal{I}}(\SNR)$ for $\nT=\nR=1$ and for $\nT=2$, $\nR=1$, with $\nb=10$. In dashed, the
  respective perfect-CSI capacities.}
  \label{MI_2x1_10}
\end{figure}

In this case, the relative gap between the perfect-CSI capacity and the achievable mutual information is very substantial.
(At $0$ dB, less than half the perfect-CSI capacity can actually be achieved by IID complex Gaussian inputs.)
The spectral efficiency of pilot-based schemes is similarly affected.
Remarkably though, the performance of these schemes relative to the mutual information limit is essentially unaffected.

\begin{example}
\label{ibra}
Let $\nR=1$ and let $\nb=10$. Shown in Fig. \ref{MI_2x1_10} is the mutual information as function of $\SNR$ with $\nT=1$ and with $\nT=2$.
Also shown are the respective perfect-CSI capacities.
\end{example}

A well-known feature of the perfect-CSI capacity is that it always increases with additional antennas, be it at the transmitter or at the receiver.
Example \ref{ibra} reflects this increase.
However, the considerations in Section \ref{baselines}, inspired by \cite{hassibi,Zheng02}, suggest that in reality activating $\nT > \nR$ antennas
would be detrimental at high $\SNR$.
Example \ref{ibra} confirms that this is indeed the case, not only at high $\SNR$ but at every $\SNR$ of interest, when the level of mobility is sufficiently high.
(For this particular example, up to around $\nb \approx 40$.)

\begin{figure}
  \begin{center}
  \includegraphics[width=3.5in]{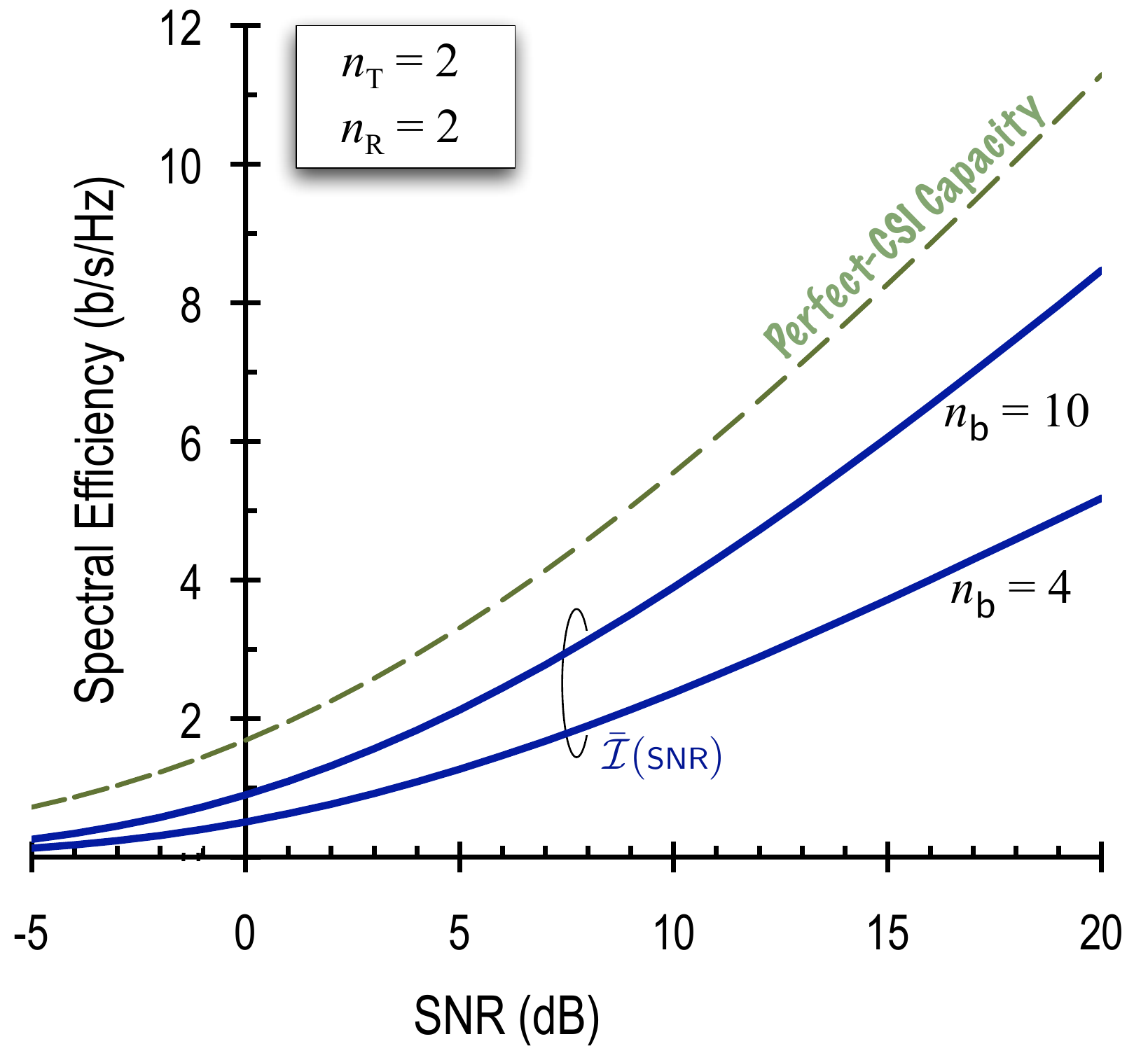}
     \end{center}
  \caption{In solid, mutual information $\bar{\mathcal{I}}(\SNR)$ for $\nT=\nR=2$, with $\nb=10$ and with $\nb=4$. In dashed, the
  corresponding perfect-CSI capacity.}
  \label{MI_2x2}
\end{figure}

\begin{example}
\label{jaime}
Let $\nT=\nR=2$. Shown in Fig. \ref{MI_2x2} is the mutual information as function of $\SNR$ with $\nb=10$
and with $\nb=4$.
Also shown is the corresponding perfect-CSI capacity.
\end{example}

Comparing Example \ref{jaime} with Examples \ref{etoo} and \ref{etoo2}, notice how, at each level of mobility,
MIMO transmission suffers a more drastic loss relative to the perfect-CSI capacity.
Note that, with $\nb=4$, the mutual information for $\nT=\nR=2$ is larger thab for  $\nT=\nR=1$ (Example \ref{etoo2}).
Although Example \ref{ibra} illustrated that an additional transmit
antenna should be activated only for reasonably long $\nb$, this shows that an additional transmit-receive
pair should be activated even for short $\nb$.



%



\section{Conclusion}

We have presented a method (partly analytical, partly Monte Carlo) to compute the mutual information achieved by IID complex Gaussian inputs on
block Rayleigh-faded channels, both scalar and MIMO. This mutual information is highly relevant as it represents the highest spectral efficiency attainable
with standard Gaussian codebooks.

In prior work \cite{JLM_ISIT09}, some of the authors derived a lower bound to this mutual information (allowing but not requiring pilot symbols).
Specialized to the case of no pilots, that bound is
\be
\bar{\mathcal{I}}(\SNR) \geq C(\SNR) - \frac{\nT \nR}{\nb} \log_2 \left( 1 + \SNR \frac{\nb}{\nT} \right) .
\ee
The method presented in this paper also allows verifying the accuracy of this lower bound.

In addition, the method may be of interest to other multivariate problems involving combinations of multiplicative and additive Gaussian noise, either with respect to the mutual information or to the constituting differential entropies.

A software routine that implements the described method in Matlab can be downloaded at 
\begin{verbatim}www.ece.umn.edu/~nihar/MI_MIMO_final.m\end{verbatim}


\section*{Appendix}

\subsection{Preliminaries}

For subsequent use, we present three relevant identities.
The first one, easily verified, is
\be \label{SnowAtXmas}  
 \int_{-\infty}^{\infty}\exp\{-x^2A+xB\}\,\mathrm{d}x=\exp\left\{\frac{B^2}{4A}\right\}\sqrt{\frac{\pi}{A}}.\ee

The second one is an integral due to Itzykson and Zuber \cite{IZ0}. Given an $M\times M$
diagonal matrix $\Zm$ with diagonal elements $\zv$, an arbitrary $M\times M$ matrix $\Dm$ with
eigenvalues $\dv$, and an $M\times M$ isotropically distributed unitary random
matrix $\Um$,
\be
\label{FreshAir}
\int e^{ \tr \{ {\mbox{\boldmath $\scriptstyle U$}} {\mbox{\boldmath $\scriptstyle D$}} {\mbox{\boldmath $\scriptstyle U$}} ^{\dagger} {\mbox{\boldmath $\scriptstyle Z$}}  \}  }
 p(\Um) \,  \mathrm{d}\Um
= \prod_{m=1}^M \frac{(m-1)!\,\,\det \Em(\dv,\zv)}{\det \Vm(\dv) \, \det \Vm(\zv)}
  \ee
where the $(i,j)$th entry of the $M\times M$ matrix $\Em(\dv,\Zm)$ equals
$\exp\{d_i z_j\}$ while $\Vm(\cdot)$ denotes a Vandermonde matrix, i.e., such that
\be
\det \Vm(\dv) = \prod_{1 \leq i,j \leq M} (d_j - d_i) .
\ee

The final identity was proved in \cite{CWZ03} by Chiani, Win and
Zanella. Given two
arbitrary $M\times M$ matrices ${\bf \Psi}(\xv)$ and ${\bf \Phi}(\xv)$
with $(i,j)$th entries $\Psi_i(x_j)$ and $\Phi_i(x_j)$, respectively, and an arbitrary
function  $\xi(\cdot)$,
{\setlength\arraycolsep{.6pt}
\bea
\label{OrganizedQueues}
\int&\cdots&\int_{\mathcal{D}_{\mathrm{ord}}} \det {\bf \Psi}(\xv)\, \det {\bf
  \Phi}(\xv)\, \prod_{m=1}^M \xi(x_m)\mathrm{d}\xv \nonumber \\
&&
=\det\left(\left\{\int_{a}^b\Psi_{i}(x)\Phi_{j}(x)\xi(x)\mathrm{d}x\right\}_{i,j=1\ldots
    M}\right) \quad\quad\quad\quad
\eea}
where  the multiple integral is over the domain
$\mathcal{D}_{\mathrm{ord}}=\{b\geq x_1 \geq x_2 \geq \ldots \geq x_M
\geq a\}$.

\subsection{Proof of Proposition \ref{R0}}

Conditioned on $\Xm$, the output $\Ym$ is complex Gaussian.
Furthermore, the rows of $\Ym$ are IID conditioned on $\Xm$.
Hence, to obtain $\mathfrak{h}(\Ym|\Xm)$ it suffices to evaluate its value for an arbitrary row of $\Ym$ and then scale it by the number of rows, i.e., by $\nR$.

Let $\yv$ be an arbitrary row of $\Ym$. The conditional covariance of the $\nb$-dimensional column vector $\yv^\dagger$ equals
\be
\E \left[ \yv^\dagger \yv | \Xm \right] = \Idm + \frac{\SNR}{\nT} \Xm^\dagger \Xm
\ee
and thus
\begin{align}
\mathfrak{h}(\yv|\Xm) & = \mathfrak{h}(\yv^\dagger|\Xm) \\
 &= E \left[ \log_2 \left( (\pi e)^{\nb} \; \det \! \left(   \Idm + \frac{\SNR}{\nT} \Xm^\dagger \Xm   \right)  \right)  \right]
\end{align}
with expectation over the distribution of $\Xm$.
Factoring out the term $\nb \log_2(\pi e)$, what remains coincides with the perfect-CSI capacity of a MIMO channel, only with the role of the channel
played by $\Xm$. Since the entries of $\Xm$ are IID complex Gaussian with zero mean and unit variance, we may directly apply the closed form in \cite{shinlee} with
appropriate dimensioning. Scaling the end result by $\nR$, we convert $\mathfrak{h}(\yv|\Xm) $ into $\mathfrak{h}(\Ym|\Xm) $ as desired.

\subsection{Proof of Proposition \ref{R1}}

Define $\gamma = \SNR/\nT$. Then, denoting by $\xv_t$ the $t$th column of $\Xm$,
{\setlength\arraycolsep{.6pt} \bea
p(\Ym)&=&\E_{{\mbox{\boldmath $\scriptstyle H$}}}\left[\int p(\Ym|\Hm,\Xm) \, p(\Xm) \, \mathrm{d}\Xm\right] \\
&=&\frac{1}{(\pi\gamma)^{\nT
  \nb}}\frac{1}{\pi^{\nR \nb}}
\E_{{\mbox{\boldmath $\scriptstyle H$}}}\left[
    \prod_{t=1}^{\nb}\int\exp\left\{-\|\yv_t-\Hm\xv_t\|^2\right\}\right.\nonumber \\
&& \;\; \left. \times \, \exp\left\{-\frac{\| \xv_t \|^2}{\gamma}\right\} \mathrm{d} \xv_t \right].
\eea
Using the singular-value decomposition $\Hm=\Um\Sm\Vm^\dagger$, absorbing $\Vm$ into $\Xm$ through the variable substitution
$\tilde{\xv}_t=\Vm^\dagger \xv_t$, and assembling the diagonal entries of $\Sm \Sm^\dagger$ into $\Lv = [ \lambda_1, \ldots, \lambda_{\nR} ]$,
{\setlength\arraycolsep{0.6pt}
\bea
\label{eq0}
p(\Ym)&=&\frac{\exp\left\{ -\|\Ym\|^2\right\} }{(\pi\gamma)^{\nT
  \nb}\pi^{\nR \nb}} \,
\E_{{\mbox{\boldmath $\scriptstyle H$}}}\left[\int\prod_{t=1}^{\nb} \exp \left\{ -\tilde{\xv}_t^{\dagger}\Lm\tilde{\xv}_t\right \} \right.\nonumber
  \\
&&\;\times
  \left.\exp\!\left\{2 \, \RE\{\yv_t^{\dagger}\Um\Sm\tilde{\xv}_t\}\right\} \exp\left\{ -\frac{ \| \tilde{\xv}_t \|^2 }{\gamma}\right\} \mathrm{d}\tilde{\xv}_t \right] \\
&=& \frac{\exp\left\{ -\|\Ym\|^2\right\} }{(\pi\gamma)^{\nT
  \nb}\pi^{\nR \nb}}  \,
\E_{ {\mbox{\boldmath $\scriptstyle H$}} } \! \left[\int\prod_{t=1}^{\nb} \left(\prod_{k=1}^{\nR}
 \right. \right. \nonumber \\
&&\;\times
 \left. \exp\left\{2 \, \RE\{\yv_t^\dagger \uv_k
      {\Sigma_k}{\tilde{x}}_{t,k}\}- |\tilde{x}_{t,k}|^2 \left(\lambda_k + \frac{1}{\gamma}\right)\right\}\right)
 \nonumber \\
&&\; \times \left.
 \left(\prod_{k=\nR+1}^{\nT}
  \exp\left\{ -\frac{| \tilde{x}_{t,k}|^2 }{\gamma}\right\} \right)\mathrm{d}\tilde{x}_{t,k}\right]
\eea}

Applying (\ref{SnowAtXmas}) to each variable $\tilde{x}_{t,k}$ in (\ref{eq0}) gives
{\setlength\arraycolsep{.6pt}
\bea
p(\Ym)&=&\frac{\exp \! \left\{\!-\|\Ym\|^2\right\} }{\pi^{\nR
  \nb}} \,
\E_{{\mbox{\boldmath $\scriptstyle H$}}} \!\!\left[\!\prod_{t=1}^{\nb}
\prod_{k=1}^{\nR}\!
  \exp\!\!\left\{\!\frac{\lambda_k \! \left(\RE\{\yv_t^\dagger\uv_k\}\right)^2}{ \left(\lambda_k+\frac{1}{\gamma}\right)} \! \right\}\right.
\nonumber \\
&&\;\times\left.
\exp\left\{\frac{\lambda_k\left( \IM\{\yv_t^\dagger\uv_k\}\right)^2}{
    \left(\lambda_k+\frac{1}{\gamma}\right)} \right\}
\left(\frac{1}{\lambda_k\gamma+1}\right)\right] \\
&=&  \frac{\exp\left\{-\|\Ym\|^2\right\} }{\pi^{\nR
  \nb}} \,
\E_{{\mbox{\boldmath $\scriptstyle H$}}} \! \left[\prod_{k=1}^{\nR}
\exp\left\{\frac{\lambda_k\gamma\uv_k^\dagger\Ym\Ym^\dagger
    \uv_k}{ (\lambda_k\gamma+1)}\right\}\right. \nonumber\\
&&\;\times \left.
\left(\frac{1}{\lambda_k\gamma+1}\right)^{\nb}  \right]  .
\label{eq_1_app}
\eea}

Let $\Am(\Lv)$ be a diagonal matrix with $k$th diagonal entry
$a_k = \lambda_k\gamma/(\lambda_k\gamma+1)$. It is known that $\Um$, $\Sm$,
and $\Vm$ in the singular-value decomposition of $\Hm$ are independent random matrices and that both $\Um$ and
$\Vm$ are isotropically distributed. Therefore,
we can express
(\ref{eq_1_app}) as
{\setlength\arraycolsep{.6pt} \bea
p(\Ym)&=&\frac{\exp\left\{-\|\Ym\|^2\right\} }{\pi^{\nR
  \nb}}\int p(\Lv)
\prod_{k=1}^{\nR}\left(\frac{1}{\lambda_k\gamma+1}\right)^{\nb}
 \nonumber \\
&&\;\times  \left[ \int
p(\Um)
\exp\left\{ \mathrm{Tr} \left\{ \Am(\Lv)\tilde{\Um}^\dagger \Ym\Ym^\dagger\tilde{\Um} \right \}\right\}
\mathrm{d}\Um\right] \mathrm{d}\Lv. \nonumber \\
\label{supermessi}
\eea}
The r.h.s. of (\ref{supermessi}) is precisly the setup in (\ref{FreshAir}). Let $\av(\Lv) = [a_1, \ldots, a_{\nR}]$ contain the
diagonal elements of $\Am(\Lv)$ and let $\dv$ contain the eigenvalues of
$\Ym\Ym^\dagger$. Then,
{\setlength\arraycolsep{.6pt} \bea \label{eq_2_app}
p(\Ym)&=&\frac{\prod_{k=1}^{\nR}
    (k-1)! \; e^{-\| {\mbox{\boldmath $\scriptstyle Y$}} \|^2} }{\det \Vm(\dv)\,\pi^{\nR
  \nb}}
\int p(\Lv) \frac{\det \Em(\av(\Lv),\dv)}{\det
  \Vm(\av(\Lv))}\nonumber \\
&&\;\times  \prod_{k=1}^{\nR}\left(\frac{1}{\lambda_k\gamma+1}\right)^{\nb}\mathrm{d}\Lv
\eea}

The density distribution of the (ordered) eigenvalues in $\Lv$ equals
\be \label{pdf}
p(\Lv)=\det^2 \Vm(\Lv) \prod_{k=1}^{\nR} \frac{ e^{-\lambda_k}
\lambda_k^{\nT-\nR}}{(\nR-k)! \, (\nT-k)!}.\ee
Moreover,
\bea\label{deta}
[\det \Vm(\av(\Lv))]^{-1}& =&
  \prod_{k>\ell}\left(\frac{\lambda_k\gamma}{\lambda_k\gamma+1}-\frac{\lambda_{\ell}\gamma}{\lambda_{\ell}\gamma+1}\right)^{-1} \nonumber \\
&=&\prod_{k>\ell}
  \frac{1}{\gamma}\frac{(\lambda_k\gamma+1)(\lambda_{\ell}\gamma+1)}{\lambda_k-\lambda_{\ell}}  \nonumber \\
&=&\frac{1}{\gamma^{(\nR^2-\nR)}}\frac{\prod_{k>\ell}
  (\lambda_k\gamma+1)(\lambda_{\ell}\gamma+1)}{\det \Vm(\Lv)}  \nonumber \\
&=&\frac{1}{\gamma^{(\nR^2-\nR)}}\frac{\prod_{k=1}^{\nR} (\lambda_k\gamma+1)^{\nR-1}}{\det \Vm(\Lv)}.
\eea
Plugging (\ref{pdf}) and (\ref{deta}) into (\ref{eq_2_app}) yields
{\setlength\arraycolsep{.6pt}
\bea \label{eq_3_app}
p(\Ym)&=&\frac{ \exp\left\{-\|\Ym\|^2\right\} }{\det \Vm(\dv)\,\gamma^{(\nR^2-\nR)}\pi^{\nR
  \nb}\prod_{k=1}^{\nR}(\nT\!-\!k)!}\int\! \det \Vm (\Lv)  \nonumber \\
&&\;\times \,\det \Em(\av(\Lv),\dv)\,  \prod_{k=1}^{\nR}\frac{
  e^{-\lambda_k}
  \lambda_k^{\nT-\nR}}{(\lambda_k\gamma+1)^{\nb+1-\nR}}\mathrm{d}\Lv
\nonumber \\
&=&\frac{ \exp\left\{-\|\Ym\|^2\right\} }{\det \Vm(\dv)\,\pi^{\nR
  \nb}\prod_{k=1}^{\nR}(\nT\!-\!k)!}\int\! \det \Vm (\Lv/\gamma)  \nonumber \\
&&\;\times \,\det \Em(\av(\Lv),\dv)\,  \prod_{k=1}^{\nR}\frac{ e^{-\lambda_k} \lambda_k^{\nT-\nR}}{(\lambda_k\gamma+1)^{\nb+1-\nR}}\mathrm{d}\Lv.
\eea

The multiple integral in (\ref{eq_3_app}) is an instance of
(\ref{OrganizedQueues}). Simply identify
\begin{align}
\Phi(\Lv)&= \det \Vm(\Lv/\gamma) \spazio \\ \spazio
\Psi(\Lv)&= \det \Em(\av(\Lv),\dv)\\
\xi(x)&= \frac{e^{-x} \, x^{\nT-\nR}} {(x\gamma+1)^{\nb+1-\nR}}
\end{align}
to obtain
\be
p(\Ym)=\frac{\exp\left\{- \left \|\Ym\right\|^2  \right\}}{\,\pi^{\nR
    \nb}}\frac{\det \Zm}{\det \Vm(\dv)\, \prod_{k=1}^{\nR}(\nT-k)!}
\ee
where
\be
Z_{ij}= \! \int_0^{\infty} \!\!\! (x/\gamma)^{i-1}\exp\!\left\{\!d_j\frac{x\gamma}{x\gamma\!+\!1}\!-\!x\!\right\}\frac{x^{\nT-\nR}}{(x\gamma\!+\!1)^{\nb+1-\nR}} \, \mathrm{d}x.
\ee


\end{document}